\documentstyle[12pt,times,psfig]{article}
\newcommand{\beq}{\begin{equation}}
\newcommand{\eeq}{\end{equation}}
\newcommand{\bea}{\begin{eqnarray}}
\newcommand{\eea}{\end{eqnarray}}

\begin{document}
\noindent{
{\large\bf 
LANL Report LA-UR-98-5998 (1998)\\
}
}
Talk given at the {\em Fourth Workshop on Simulating Accelerator Radiation
Environments (SARE4)}, \\
Knoxville, Tennessee, September 14-16, 1998

\begin{center}
{\Large \bf 
Modeling Fission in the Cascade-Exciton Model}\\
\vspace*{0.3cm}
{\bf Arnold~J.~Sierk and Stepan~G.~Mashnik}\\
\vspace{0.3cm}
{\it T-2, Theoretical Division, Los Alamos National Laboratory,
Los Alamos, NM 87545}\\
\end{center}
\begin{abstract}
Recent developments of the Cascade-Exciton Model (CEM) of nuclear
reactions to describe high energy particle induced fission are
briefly described. The increased accuracy and predictive power of 
the CEM are shown by several examples.  Further necessary work is 
outlined.
\end{abstract}
\begin{center}
{\large 1. Introduction} \\
\end{center}

A number of recent projects like accelerator transmutation of waste, 
accelerator-based conversion, accelerator-driven energy production,
accelerator production of tritium, etc.~have revived interest in
reliable data from different types of nuclear reactions,
including high energy fission (see, e.g., \cite{kalmar96}).

Many efforts have been previously made to develop models of high energy 
fission and to implement them into codes used in applications. 
For example, the Fong statistical model of fission~\cite{fong}
has been incorporated into different Monte Carlo codes, e.g., at 
JINR~\cite{toneev66,shmakov}, ORNL~\cite{al81}, BNL~\cite{taka84}, 
and PINP, Gatchina~\cite{nesterov82}. A similar approach based on the
thermodynamical model of fission was developed in Moscow at ITEP
by Stepanov~\cite{stepanovfis} and was incorporated in the ITEP code
INUCL.
A more sophisticated and physically grounded, but at the same time,
more complicated and time consuming approach is the dynamical
``Diffusion Model of Fission"~\cite{gonchar} implemented as a
Monte Carlo code at INR, Moscow by Mebel et al.~\cite{mebel}. 
On the other hand, several much simpler
approaches based on phenomenological distributions of fragments, such as 
Atchinson's model~\cite{atch94}, Nakahara's model~\cite{nakahara},
and Rubchenya's recent model~\cite{rubchenya} are also
used in a number of codes. Finally, in a number of
current works, high energy fission and fragmentation is calculated
in a single statistical model of sequential binary decays, using
the code GEMINI~\cite{gemini}.

Fission is a slow process, which at intermediate and high incident energies 
is considered to take place after fast processes, described usually
by intranuclear cascade models (INC), intermediate or preequilibrium
processes, and evaporation from compound nuclei before fission. One
of the models with a good predictive power for the first three stages
of reactions preceding fission is the Cascade-Exciton Model (CEM)
of nuclear reactions \cite{cem} in its currently improved modifications
(see \cite{cem98} and references therein). 
CEM, in its standard
version \cite{cem}, did not consider fission at all. Lately, the model
has been extended by taking into account the competition between
particle emission and fission at the compound nuclear stage
and a more realistic calculation of nuclear level density \cite{acta2}.
A version of the CEM with these modifications, as realized in the code
CEM92, was used succesfully by Konshin \cite{konshin} to calculate
nucleon-induced fission cross sections for actinides in the
energy region from 100 MeV to 1 GeV. Neveretheless, both CEM92 and a later
version of the code, CEM95 \cite{cem95}, run into problems when
used for preactinides \cite{prokofiev}. In addition, CEM95 only 
allows us to calculate nuclear fissilities and fission cross sections 
but not the process of fission itself, and does
not provide fission fragments and a further possible evaporation of
particles from them. When, during the Monte Carlo simulation of the
compound stage of a reaction with CEM95, 
we encounter a fission, we simply remember this event (that 
permits us to calculate fission cross section and fissility)
and finish the calculation of this event 
without a subsequent calculation of fission fragments and a further 
possible evaporation of particles from them.

This means that CEM95 is suitable for calculating yields
of produced nuclides only in the spallation region. To extend the
range of its applicability into the fission and fragmentation regions,
it should be developed further. Our so far modest progress in 
modeling fission with the CEM is described in this paper.\\

\begin{center}
{\large 2. Modeling Fission Cross Sections} \\
\end{center}

Because we consider in detail only the third, compound-decay stage of
nucleon-induced fission reactions, we will not discuss here the details of
the modeling of the cascade and preequilibrium stages of the reactions
\cite{cem,cem98,cem95}. We give below a brief summary of the models
used in calculating fission and particle decay widths, and define the 
parameters whose variation is discussed in Section 3.

We approximate the partial widths $\Gamma _j$ for the emission of a particle $j$
($j \equiv $ n, p, d, t, $^3$He, $^4$He) and $\Gamma _f$ for fission by the
expressions:

\begin{equation}
\Gamma _j = {{(2s_j + 1) m_j} \over {\pi ^2 \rho_c (U_c)}}
\int \limits_{V_j}^{U_j - B_j} \sigma_{inv}^j (E) \rho_j (U_j - B_j - E)
E dE \mbox{ ,}
\end{equation}

\begin{equation}
\Gamma _f = {1 \over {2 \pi \rho_c (U_c)}}
\int \limits_0^{U_f - B_f} \rho_f (U_f - B_f - E) dE \mbox{ ;}\\
\end{equation}
where $\rho _c$, $\rho _j$, and $\rho _f$ are the level densities of the 
compound nucleus, the residual nucleus produced after the emission of
the $j$-th particle, and of the fissioning nucleus at the fission saddle 
point, respectively; $m_j$, $s_j$ and $B_j$ are the mass, spin and the 
binding energy of the $j$-th particle, respectively, and $B_f$ is the 
fission-barrier height. $\sigma_{inv}^j (E)$ is the
inverse cross-section for absorption of the $j$-th
particle with kinetic energy $E$ by the residual nucleus.  The ``thermal"
energies $U_k$ are defined by 
$$U_c = E^* - E_R^c - \Delta_c \mbox{ ; } U_j = E^* - E_R^j - \Delta_j
 \mbox{ ; } U_f = E^* - E_R^{sp} - \Delta_f\mbox{ ,}$$\
where $E^*$ is the total excitation energy of the compound nucleus, and
$E_R^c$ and $E_R^j$ are the rotational energies of the compound and
residual nuclei at their ground state deformations
$$E_R^{gs} = {L(L+1)\hbar ^2\over 2J_{rb}}\mbox{ ,}$$\
where the rigid-body moment of inertia is approximated as
$$J_{rb} = 0.4 m_N r_0^2 A^{5/3} \mbox{ .}$$
For the fission saddle point, 
$$E_R^{sp} = {L(L+1)\hbar ^2\over 2J_{sp}} \mbox{ ,}$$\\
and $J_{sp}$ , the moment of inertia of the saddle-point shape, is taken
from Ref.~\cite{strutinsky} or \cite{cohen}.

Following Ref.~\cite{iljinov92}, the pairing energies of the compound
nucleus~$\Delta_c$, of the residual nucleus~$\Delta_j$, and
of the fission saddle point~$\Delta_f$ are estimated as:
\begin{equation}
\Delta_c = \chi_c \cdot 12 / \sqrt{A_c} \mbox{ ; }
\Delta_j = \chi_j \cdot 12 / \sqrt{A_{fj}} \mbox{ ; and }
\Delta_f = \chi_c \cdot 14 / \sqrt{A_c} \mbox{ (in MeV).}
\end{equation}
$A_{fj} = A_c - A_j$,  where $A_c$ and $A_j$ are the
mass numbers of the compound nucleus and of the $j$-th particle, 
and $\chi_k = 0$, 1, or 2 for odd-odd, odd-even, or even-even nuclei,
respectively.  For the inverse cross sections, the approximations proposed 
by Dostrovsky et al.~\cite{dostrovsky} are used,
\begin{equation}
\sigma_{inv}^j (E) = \sigma_{geom}^j \alpha _j \left(
1 + {\beta_j \over E} \right) \mbox{ ,}
\end{equation}
where 
\begin{equation}
\sigma_{geom}^j = \pi R_j^2 \mbox{ ; } R_j = \tilde r_0 A_{fj}^{1/3}
 \mbox{ ; } \tilde r_0 = 1.5 \mbox{ fm ;}
\end{equation}
$$\alpha_n = 0.76 + 2.2 A_{fj}^{-1/3} \mbox{ ;}$$
$$\beta_n = (2.12 A_{fj}^{-2/3} -0.05) / \alpha_n \mbox{ .}$$\
For charged particles $\beta _j = - V_j$ , where $V_j$ is the effective
Coulomb barrier and the constants $\alpha_j$ are calculated for
a given nucleus by interpolating the values of Ref.~\cite{dostrovsky}.

For the level density, we use the simple form 
$$\rho(U_{eff}) \simeq Const \cdot exp\{ 2 \sqrt{aU_{eff}} \} \mbox{ ,}$$
where $U_{eff}$ is the argument of $\rho$ in Eqs.~(1) and (2).
For the level density parameter $a_j$, we consider either a constant,
one empirical determination of $a_j(Z,N)$ \cite{malyshev}, or eight different 
sets of parameters of the form $a_j(Z,N,E^*)$, each with three empirically 
determined parameters\cite{iljinov92}, \cite{ignatyuk1}--\cite{cherepanov80}.
\begin{equation}
a(Z,N,E^*) = \tilde a(A) \left\{ 1 + \delta W_{gs}(Z,N)
{{f(E^* - \Delta)} \over {E^* - \Delta}} \right\},
\end{equation}
where
\begin{equation}
\tilde a(A) = \alpha A + \beta A^{2/3} B_s 
\end{equation}
is the asymptotic Fermi-gas value of the level density parameter at
high excitation energies.
Here, $B_s$ is the ratio of the surface area of the nucleus to the
surface area of a sphere of the same volume (for the ground state of a
nucleus, $B_s \approx 1$), and
\begin{equation}
f(E) = 1 - exp (-\gamma E) \mbox{ .}
\end{equation}
The shell and pairing corrections $\delta W_{gs}(Z,N)$ (which also enter
the calculation of the binding energies $B_j$) may be approximated using
the systematics due to Cameron~\cite{cameron57},
Truran, Cameron and Hilf~\cite{cameron70}, or 
Myers and Swiatecki~\cite{myersswiat}.  The fission barriers $B_f$ are
calculated by one of several macroscopic models \cite{barashenkov73}--
\cite{s86}, with the addition of $\delta W_{gs}(Z,N)$.  In addition, we may
use a temperature-dependent barrier~\cite{barashenkov74,sauer} or one 
independent of temperature.
The level density parameter at the saddle point, $a_f$, used in Eq.~(2), 
is assumed to be a constant times the value of $a_n$.

As the fission cross sections for preactinides constitute a 
small part of the total reaction cross section, the statistical weight
method~\cite{barashenkov74}
is used to calculate fission cross sections instead of the
straightforward Monte-Carlo technique. It allows us to obtain statistical
uncertainties of the calculated fission cross sections generally of
not more than 5--10\% with only 3000 inelastic events in each calculation.
Following Ref.~\cite{barashenkov74} we use the statistical functions
$W_n = \prod _{i=1}^N w_{ni}$ and $W_f = 1 - W_n$.
Here, $W_n$ is the
probability of the nucleus to ``drop" the excitation energy $E^*$ by a
chain of $N$ successive evaporations of particles; $W_f$ is the
probability for the nucleus to fission at any of the chain stages;
$w_{ni} = 1 - w_{fi}$ is the probability of particle emission at the
$i$-th stage of the evaporative process; $w_{fi}$ is the corresponding
fission probability which is easy to determine using the formulae 
(1,2) for the widths $\Gamma_j$ and $\Gamma_f$. After the subsequent 
averaging of $W_f$ over the total number $N_{in}$ of the cascades
followed, and after multiplication of the result by the corresponding
total inelastic cross section $\sigma_{in}$, we obtain
the following expression for the fission cross section:
\begin{equation}
\sigma_f = {{\sigma_{in}}
\over {N_{in}}}
\sum_{i=1}^{N_{in}} (W_f)_i \mbox{ .}
\end{equation}

This approch is used in a version of the CEM as realized in the code 
CEM95 \cite{cem95}. It allows us to calculate quite reliable fission
cross sections for actinides in a large range of incident energies, 
provided the level density papameter at the saddle point $a_f$, 
or more exactly, the ratio $a_f /a_n$ can be varied to reproduce data. 
As an example, the incident energy dependence of experimental
\cite{li91,ei94} and calculated  fission
cross section for interaction of neutrons with $^{238}$U is shown
in Fig.~1. \\


\begin{figure}[h!]
\centerline{
\psfig{figure=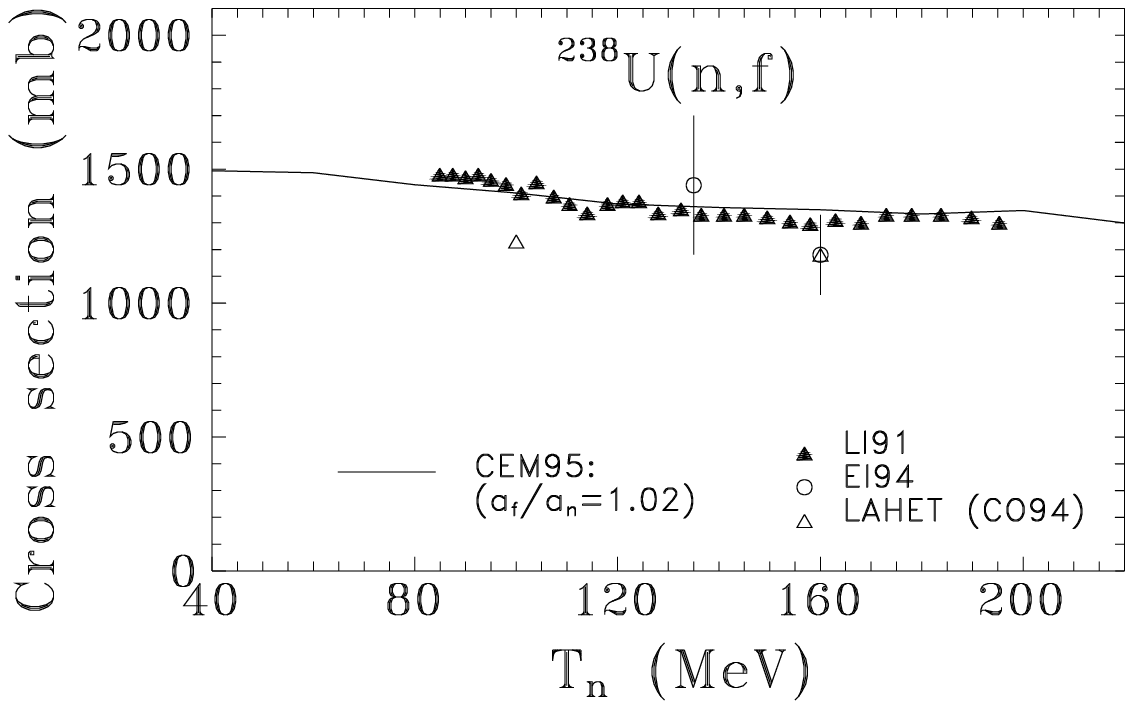,width=170mm,angle=0}\hspace{+5mm}}
\end{figure}

\vspace{-0.2cm}

{\small 
Fig.~1. Energy dependence of the neutron-induced fission cross section
of $^{238}$U. Calculations are performed with Krappe, Nix, and Sierk
fission barriers~\cite{kns}, Cameron shell and pairing corrections
\cite{cameron57}, the third Iljinov et al.~systematics for the level
density parameters~\cite{iljinov92}, with a dependence $B_f(L)$ estimated
by a phenomenological approach (formulas (28--30) from Ref.~\cite{acta2})
with the value for the moment of inertia of a nucleus at the saddle-point
$J_{sp}$ from Ref.~\cite{strutinsky}, without taking into account the
dependence of $B_f$ on excitation energy $E^*$, and with the value for
the ratio $a_f/a_n = 1.02$. The experimental points are from 
Refs.~\cite{li91,ei94}. For comparison, two values
of $\sigma_f$ calculated at $T_n = 100$ and 160 MeV in Ref.~\cite{co94}
with the code LAHET~\cite{lahet}  are shown by open triangles.}   

\noindent{
We performed these calculations with Krappe, Nix, and Sierk
fission barriers~\cite{kns}, Cameron shell and pairing corrections
\cite{cameron57}, the third Iljinov et al.~systematics for the level
density parameters~\cite{iljinov92}, with a dependence $B_f(L)$ estimated
by a phenomenological approach (formulas (28--30) from Ref.~\cite{acta2})
with the value of the moment of inertia of a nucleus at the saddle-point
$J_{sp}$ from Ref.~\cite{strutinsky}, without taking into account the
dependence of $B_f$ on excitation energy $E^*$, and with the value for
the ratio $a_f/a_n = 1.02$. For comparison, two values of fission cross
sections calculated in Ref.~\cite{co94} with the well known code
LAHET \cite{lahet} are shown for $T_n = 100$ and 160 MeV. One can see
that CEM95 describes correctly (and as well as LAHET) the shape
and the absolute value of the fission cross section at these intermediate
incident energies with this choice for $a_f /a_n$.}

\begin{center}
{\large 3. Further Development of the Model} \\
\end{center}

Our analysis
\cite{acta2,prokofiev} shows that fission cross sections calculated 
with CEM95 both for actinides and preactinides
are the most sensitive to the ratio of the level density parameters 
in the fission and neutron emission channels, $a_{f}/a_{n}$.
The variation of the $a_{f}/a_{n}$ value by only a few percent from the
optimal value makes
the cross sections vary by a factor of 1.5--2, although the
shape of the excitation function does not change appreciably.
The sensitivity of calculated fission cross sections to other CEM95
input parameters, i.e., to the choice for the level density parameter 
systematics, nuclear masses, shell and pairing corrections, saddle-point
moments of inertia, and for the macroscopic and microscopic fission 
barriers is lower. The majority of these choices give nearly the same
shape to the fission excitation function. The absolute scale of the 
excitation function varies from one model to another. For any model
it is possible to improve the agreement with experiment by adjusting
the $a_{f}/a_{n}$  parameter. However, it is not possible to alter the 
shape of the calculated excitation functions, which for preactinides
are typically too steep above 100 MeV \cite{prokofiev}.

Since CEM95 allows us to describe well the characteristics of nuclear 
reactions that do not involve fission, like double differential spectra of
secondary particles as well as spallation product yields (see \cite{cem98} 
and references therein), it is natural to search for the reasons
for the discrepancies in the modeling of fission cross sections for
preactinides observed in \cite{prokofiev}.
Two different possibilities are:

1) taking into account dynamical effects in fission, which reflect the
connection
between single-particle and collective nuclear degrees of freedom (see, for
example,
\cite{weid89}). 
The diffusive character of nuclear motion towards
and over the saddle point due to nuclear viscosity leads to a decrease of the
value of the fission width. This effect may rise with the excitation
energy, i.e., in just the direction needed for a better description of the
experimental fission cross sections \cite{prokofiev};

2) modification of the calculation of the level density at the saddle
point. This possibility \cite{prokofiev} is discussed below. 

As mentioned above, CEM95 incorporates several different level density
parameter systematics. Most of them utilize formula (6) for the excitation
energy dependence of the level density parameter, 
which was suggested first by Ignatyuk et al.~\cite{ignatyuk2}.
It reflects the effect of the strong correlation between the single-particle 
state density and the shell correction magnitude for low excitation energies,
and the fade out of the shell effects on the level density for high excitations
\cite{ignatyuk1,ignatyuk2}.

In the original version of CEM95, the level density parameter systematics 
based on formulae (6) and (7) are applied for all decay channels of an excited
nucleus except for the fission channel. In the latter case, the level density
parameter at the saddle point $a_{f}$ is calculated using an
analogous parameter for the neutron emission channel, $a_{n}$, and a constant
ratio, $a_{f}/a_{n}$, which serves as a fitting parameter of the model. Thus
the shell-effect influence on the level density in the
neutron emission channel is automatically conveyed to the level density at the
saddle point. On the other hand, we expect that shell corrections at the
saddle point should bear no relation to those at
the ground state, due to the large saddle-point deformation, and the
consequent different microscopic level structure near the Fermi surface.
In fact, the shell corrections at the saddle point should
be of much smaller magnitude than those at the ground state
for nuclei with a spherical ground-state shape,
due to the greatly reduced symmetry
of the saddle-point shape compared to the ground state
(see also Refs.~\cite{iljinov80,giardina94}).

In order to estimate the importance of taking into account this effect, 
we perform calculations for neutron- and proton-induced reactions on 
$^{208}$Pb and $^{209}$Bi with the parameter $a_{f}$ being
{\it energy-independent}, which is equivalent to the disappearance
of the shell-effect influence on the level density
at the saddle point. The magnitude of the parameter $a_{f}$
is calculated using formula (7) with the same values of the coefficients
$\alpha$ and $\beta$ that are utilized for the other decay channels. 
The parameter $B_{s}$ is adjusted
to provide the best agreement with the experimental fission cross
section approximations. All the other model parameters  are fixed and the
same as for the calculations described in Ref.~\cite{prokofiev}.
The optimal values of the parameter $B_{s}$  are
1.18 and 1.15 for interactions of protons with $^{209}$Bi and $^{208}$Pb,
and 1.12 for neutron-induced reactions for both $^{209}$Bi and $^{208}$Pb,
respectively.

The values for the parameter $B_s$
are physically reasonable because they are larger than 1, which reflects the
larger deformation of the fissioning nucleus at the saddle point in comparison
with the equilibrium state and they are not far from the value
$B_{s} \approx 2^{1/3}$, which is expected for the saddle-point configuration 
of the preactinides \cite{ignatyuk85}. 
However, we do not yet understand why different
values are needed for proton- and neutron-induced fission, and why the values
of $B_{s}$ for incident neutrons are significantly smaller than those
corresponding to the known deformation of saddle-point shapes in this
mass region.

The ratio $a_{f} / \widetilde{a_n}$, where $\widetilde{a_n}$ is the
asymptotic (large $E^*$) level density parameter for
neutron emission, is fixed by the fitting of B$_{s}$, and is not independently
fitted as has to be done in the standard CEM95. $\widetilde{a_n}$ is calculated
using Eq.~(7) and the values $\alpha=0.072$ and $\beta=0.257$,
which correspond to the 3rd Iljinov et al.~systematics \cite{iljinov92}.
The ratios  appear to depend only weakly on the nuclear mass. For
the reactions under study their values (given in Fig.~2) are in the range
1.045--1.070.

The results of calculations with the modified CEM95 are shown in Fig.~2.
We see a much better description of the experimental
fission cross sections in comparison with the original version 
for nucleon energies of 100--500 MeV. On the
other hand, the calculation systematically overestimates
to a slight extent the fission cross
sections below 100 MeV. We emphasize that other related improvements
need to be made before a final assessment of the model's value and
predictive capability can be made.  For example, an expected
excitation-energy dependence of the ground-state shell correction 
would change the average height of the calculated fission barriers as 
the incident energy is increased. 

To test these modifications to CEM95 for other types 
of nuclear reactions, we performed calculations for a number of
photon- and pion-induced reactions as well. 
One example is shown in Fig.~3.
We see that the modified version of CEM95 allow us to get a much 
better agreement with the experimental data also for photofission 
reactions. Similar improvements were obtained for 
fission cross sections induced by intermediate energy $\pi^-$
on Sn, on Bi, as well as on the actinide $^{238}$U nucleus \cite{peterson98}.

Although these results are only preliminary and questions still
remain, e.g., about the optimal value of the parameter $B_s$;
the current modifications \cite{prokofiev} do indicate the crucial importance
of properly incorporating appropriate level densities and motivate the
search for a consistent model of barriers, ground-state masses, and level
densities which may improve the predictive power of the CEM.
Besides this modification of the CEM95 code introduced especially 
for a better description of fission cross sections, we have been working 
on further improvement to the CEM \cite{cem98}, striving for a 
model capable of predicting different characteristics of nuclear reactions 
for arbitrary targets in a wide range of incident energies. 
Many of the modifications made for a better description of the
preequilibrium, evaporative, and even for the cascade stages of reactions
will affect as well the fission channel.
So, we have incorporated into the CEM the 

\newpage
\begin{figure}[h!]
\centerline{
\psfig{figure=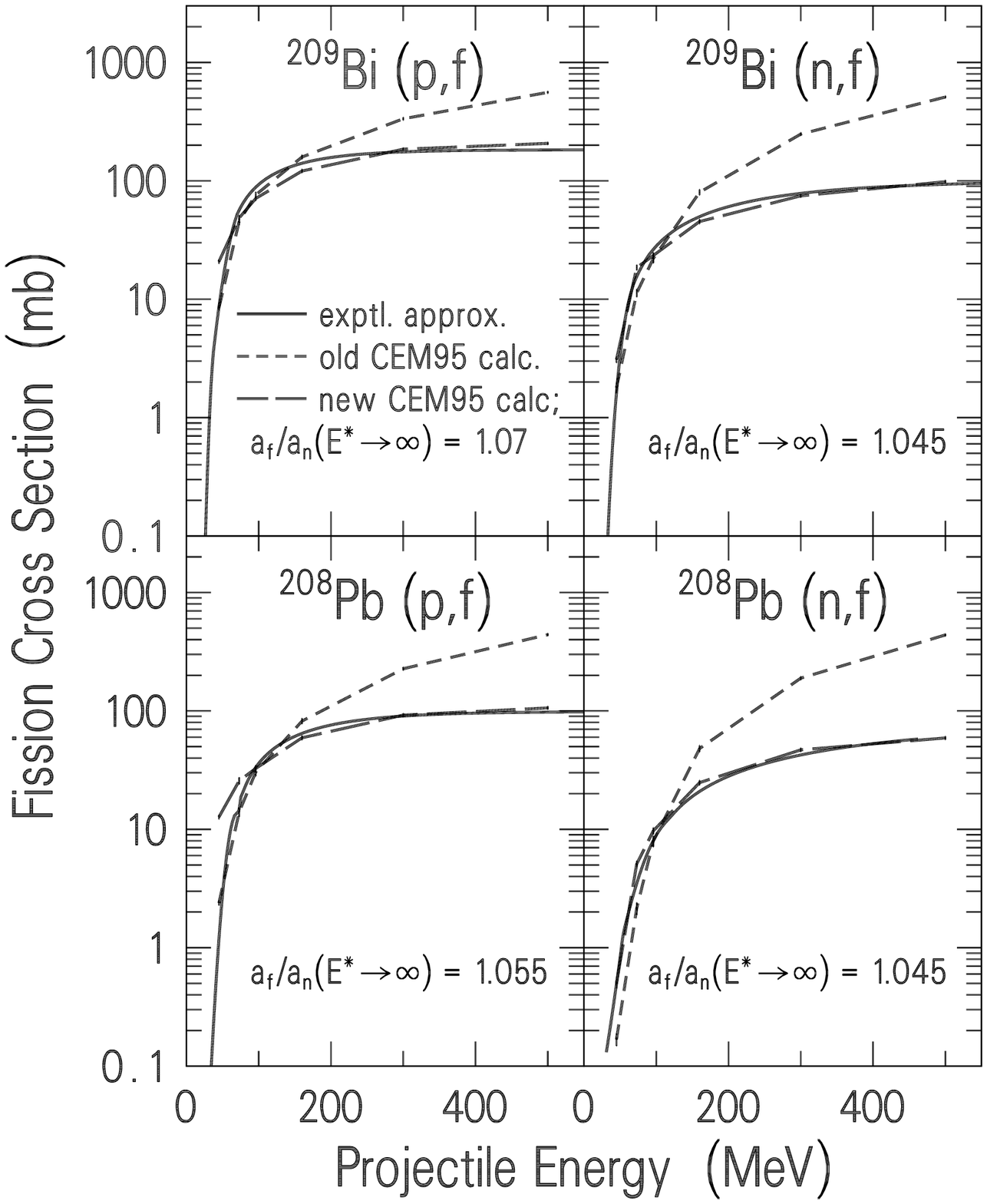,width=165mm,angle=-0}\hspace{+5mm}}
\end{figure}
\vspace*{-0.1cm}

{\small
Fig.~2.
Comparison between the experimental data                
and calculations of the cross sections for the reactions
$^{209}$Bi(p,f), $^{209}$Bi(n,f), $^{208}$Pb(p,f),
and $^{208}$Pb(n,f) using the original and modified
versions of CEM95.
The solid lines represent the approximation of the experimental
data according to \cite{prokofiev}.
The short-dashed lines show the original CEM95 results
and the long-dashed lines, the results after the modification 
of the fission channel in CEM95.
}

\newpage

\begin{figure}[h!]
\vspace*{-3.5cm}
\centerline{
\psfig{figure=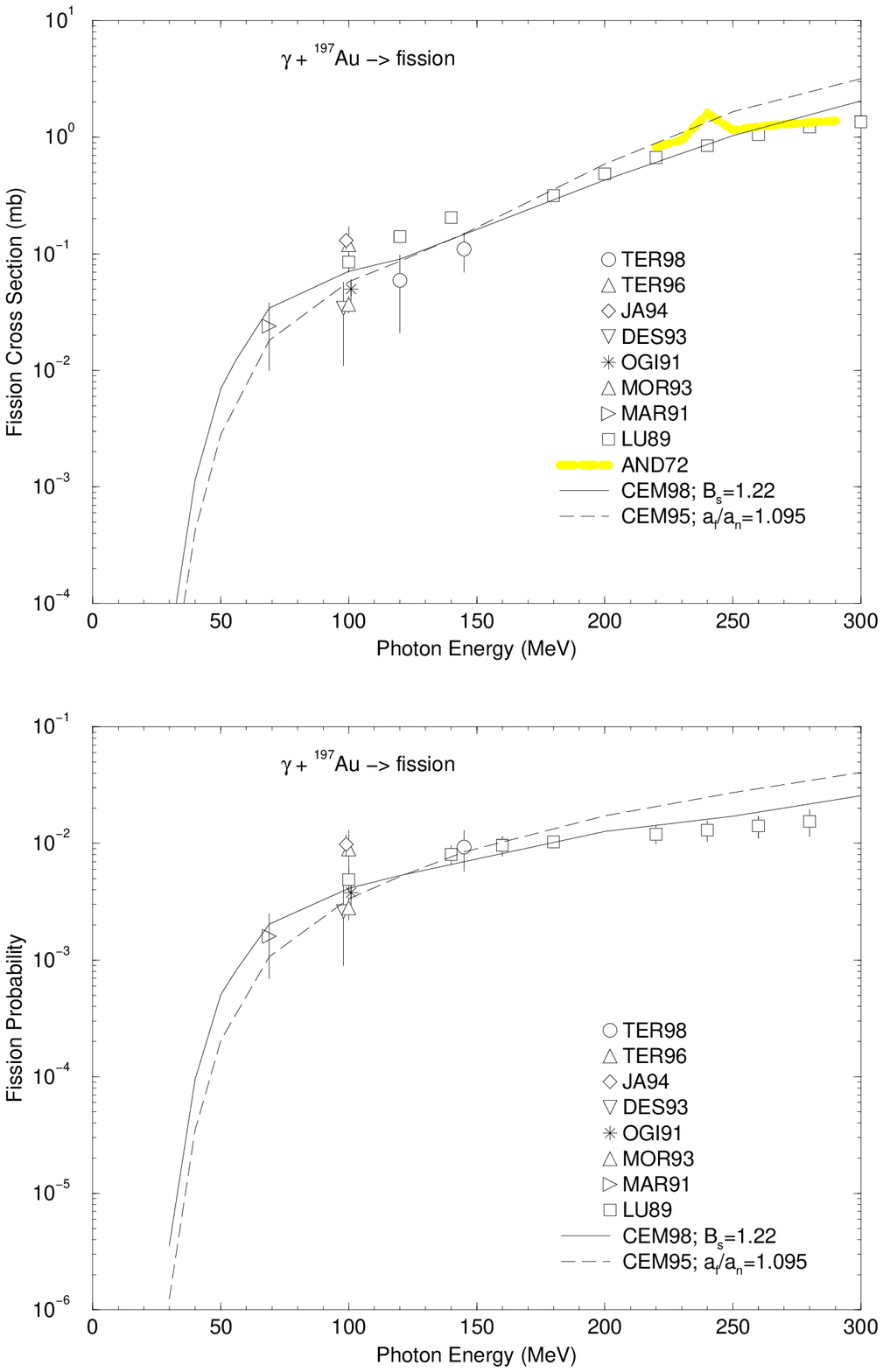,width=200mm,angle=-0}\hspace{+5mm}}
\end{figure}
\vspace*{-3.3cm}

{\small
Fig.~3.
Comparison between the experimental data \cite{ter98}--\cite{and72}
and the calculations on fission probability and fission cross section for 
photofission of $^{197}$Au using the original and modified (noted in this
figure as CEM98) versions of the CEM95 code.\\
}

\noindent{
updated experimental atomic
mass table by Audi and Wapstra \cite{audi93},
the nuclear ground-state masses,  deformations, and shell corrections by 
M\"oller et al.~\cite{moller95}, and
the pairing energy shifts from M\"oller, Nix, and Kratz \cite{moller97} 
into the level density formula. In addition, we have derived a corrected
systematics for the level density parameters using the Ignatyuk
expression Eqs.~(6,7), with coefficients fitted to the data analyzed
by Iljinov et al.~\cite{iljinov92} (we discovered that Iljinov
et al.~used $11 / \sqrt{A}$ for the pairing energies $\Delta$ (see Eq.~(3))
in deriving their level density systematics instead of the value of  
$12 / \sqrt{A}$ stated in Ref.~ \cite{iljinov92} and found several misprints 
in nuclear level density data shown in their Tabs.~1 and 2 used in fitting). 
We derived also additional semiempirical level density parameter systematics 
using the M\"oller et al.~\cite{moller95}
ground state microscopic corrections, both with and without
the M\"oller, Nix, and Kratz \cite{moller97} pairing gaps. We also introduced 
in the CEM a new empirical relation to take into account the 
excitation-energy dependence of
the ground-state shell correction $\delta W_{gs}$ in the calculation 
of fission barriers:
\begin{equation}
B_f(A,Z,L,E^*) = B_f^0(A,Z,L) 
- \delta W_{gs}  \times \exp{(-0.3 \sqrt{E^*})} \mbox{ ,}
\end{equation}
\noindent{
where $B_f^0(A,Z,L)$ is the macroscopic fission barrier of a 
nucleus with $(A,Z)$ and an angular momentum $L$ calculated with 
the {\bf BARFIT} routine\cite{s86}, $\delta W_{gs}$ is the ground 
state shell correction by M\"oller, Nix, Myers, and Swiatecki \cite{moller95},
and $E^*$ is the excitation energy with the rotational energy removed.
The excitation energy dependence, $\exp{(-0.3 \sqrt{E^*})}$, was
derived here empirically for a better description of fission cross sections.
We incorporate also calculation of the pairing gap at the saddle point 
according to  M\"oller, Nix, Myers, and Swiatecki \cite{moller95}:
\begin{equation}
\Delta_{fiss} = 4.8 B_s  \times
[1/ (Z^{1/3}) \mbox{ (if Z even) } +  
1/ (N^{1/3}) \mbox{ (if N even}) ]\mbox{ .} 
\end{equation}

After incorporation of all these improvements into CEM95 we find a much
better description for a number of nucleon-induced fission cross sections.
As an example, Fig.~4 shows neutron-induced fission cross section of gold
recently measured by Staples \cite{staples} compared with standard CEM95 
calculations and with the improved version described above, for which $a_f/a_n$ 
and the value of the coefficient of  $\sqrt{E^*}$ were the parameters varied. 
The former results were obtained without taking into account shell corrections 
in the calculation of the level density at the saddle point, i.e., using Eq.~(7)
as discussed in the beginning of this section but without an adjustment of 
the value for $B_s$ as we used for the results shown in Figs.~2
and 3; instead we used the CEM95 default value of $B_s = 1.0$. One can see 
that the improved model reproduces the data very well, while the standard 
CEM95 has big problems for this reaction.

Our work is not finished. For instance, we are not satisfied with
the situation that in the improved versions of the CEM we still have
an additional 
input parameter to describe fission cross sections: either $B_s$,
in the approach \cite{prokofiev} illustrated in Figs.~2 and 3, or $a_f/a_n$, 
in our later version shown in Fig.~4. Of course, it is possible to fit these 
input parameters to available fission cross sections, then to 
incorporate the fit into the code, as was done, e.g., in Atchinson's model 
\cite{atch94} used in LAHET \cite{lahet} or Stepanov's model \cite{stepanovfis}
used in the INUCL code at ITEP and in a number of other similar codes.  
But such an approach seems to be lacking in predictive power for 
unmeasured reactions, and we hope to find a better solution to the problem.

Concerning modeling of the fission processes themselves, i.e.,
description of the Z, A, angle and energy distributions of fragments with
a possible further evaporation from (or even fission of) fragments, we are
just at the very beginning of the work. As a ``zeroth-order approach" to the 
problem, we attempted so far only to use in CEM95 after the evaporation stage 
of a reaction, when we have to simulate a fission,
the well-known code GEMINI \cite{gemini} and Stepanov's 
model \cite{stepanovfis} in their original versions.   
Our first results are inadequate. This is not surprising, as the
distributions of residual nuclei after the evaporation
stages of reactions, before fissioning, with respect to the 
mass $A$, charge $Z$, excitation energy $E^*$, momentum $P$, and
angular momentum $L$ calculated with the CEM
will not be the same as those calculated with INUCL which succesfully
uses Stepanov's fission model \cite{stepanovfis}, or with those from
the Li\'ege INC by Cugnon et

\newpage

\begin{figure}[h!]
\vspace*{1.0cm}
\centerline{
\psfig{figure=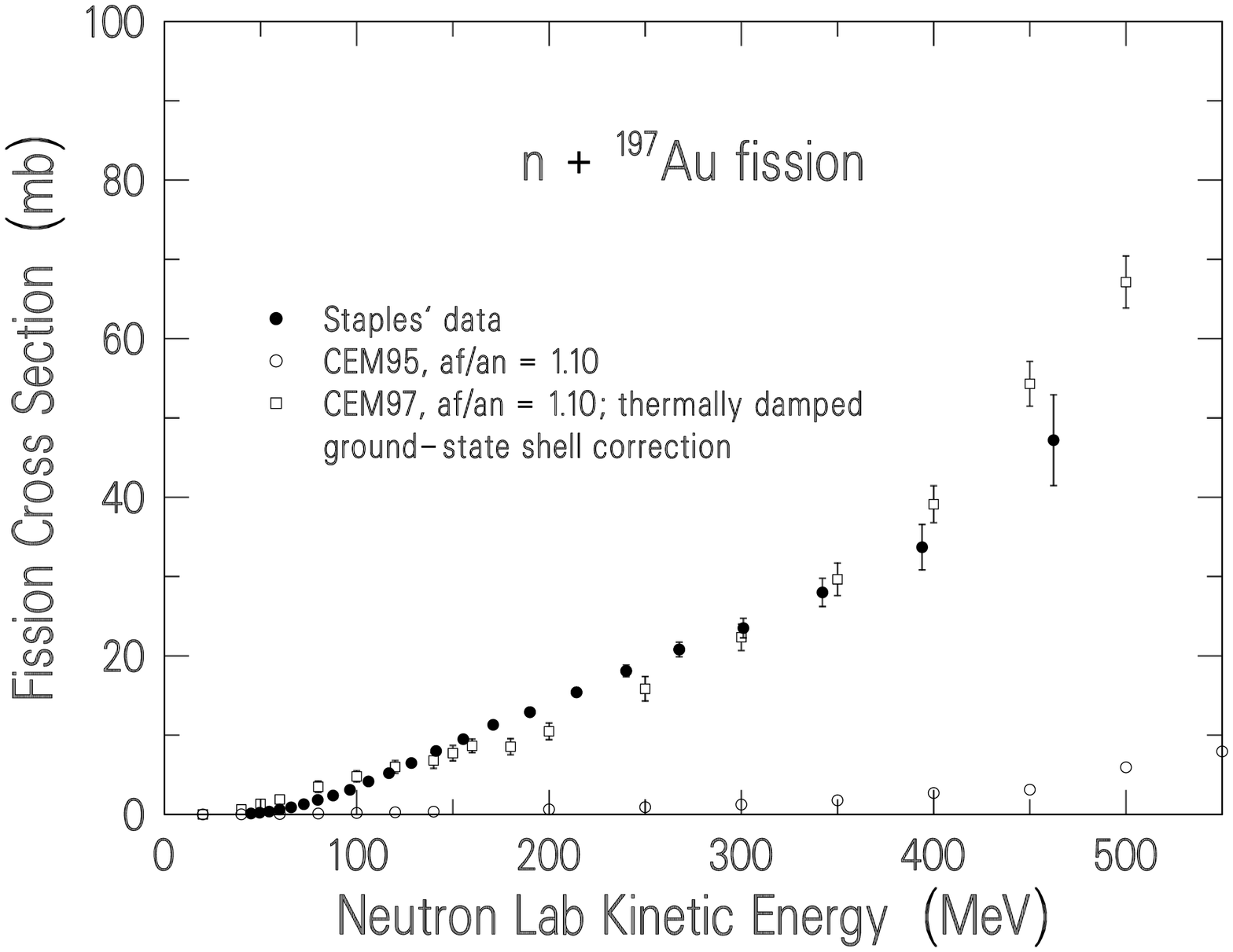,width=190mm,angle=-90}}
\end{figure}
\vspace*{-0.1cm}

{\small
Fig.~4.
The energy dependence of the neutron-induced fission cross section of gold.
Experimental data (filled circles) are by Staples \cite{staples}. CEM95
calculations (opened circles) are perfomed with fission barriers from \cite{kns},
shell and pairing corrections from \cite{cameron70},
level-density-parameter systematics from ~\cite{iljinov92}, no 
dependence of $B_f$ on $E^*$, and $a_f/a_n = 1.10$. The new calculations 
(open squares, noted in this figure as CEM97) are made with ground-state 
masses and shell corrections from \cite{moller95}, pairing energy shifts 
in the level density formula from \cite{moller97},
level density systematics corrected as described in the text in the
Ignatyuk form with coefficients fitted to the data compiled 
by Iljinov et al.~\cite{iljinov92}, macroscopic fission barriers from 
the BARFIT routine \cite{s86}, the microscopic parts of fission barriers with
thermally damped ground-state shell corrections according Eq.~(10),
the level density at the saddle point without shell corrections, and the
pairing gap at the saddle point from \cite{moller95}, cf.~Eq.~(11).\\
}

\newpage
\begin{figure}[h!]
\vspace*{0.6cm}
\centerline{
\psfig{figure=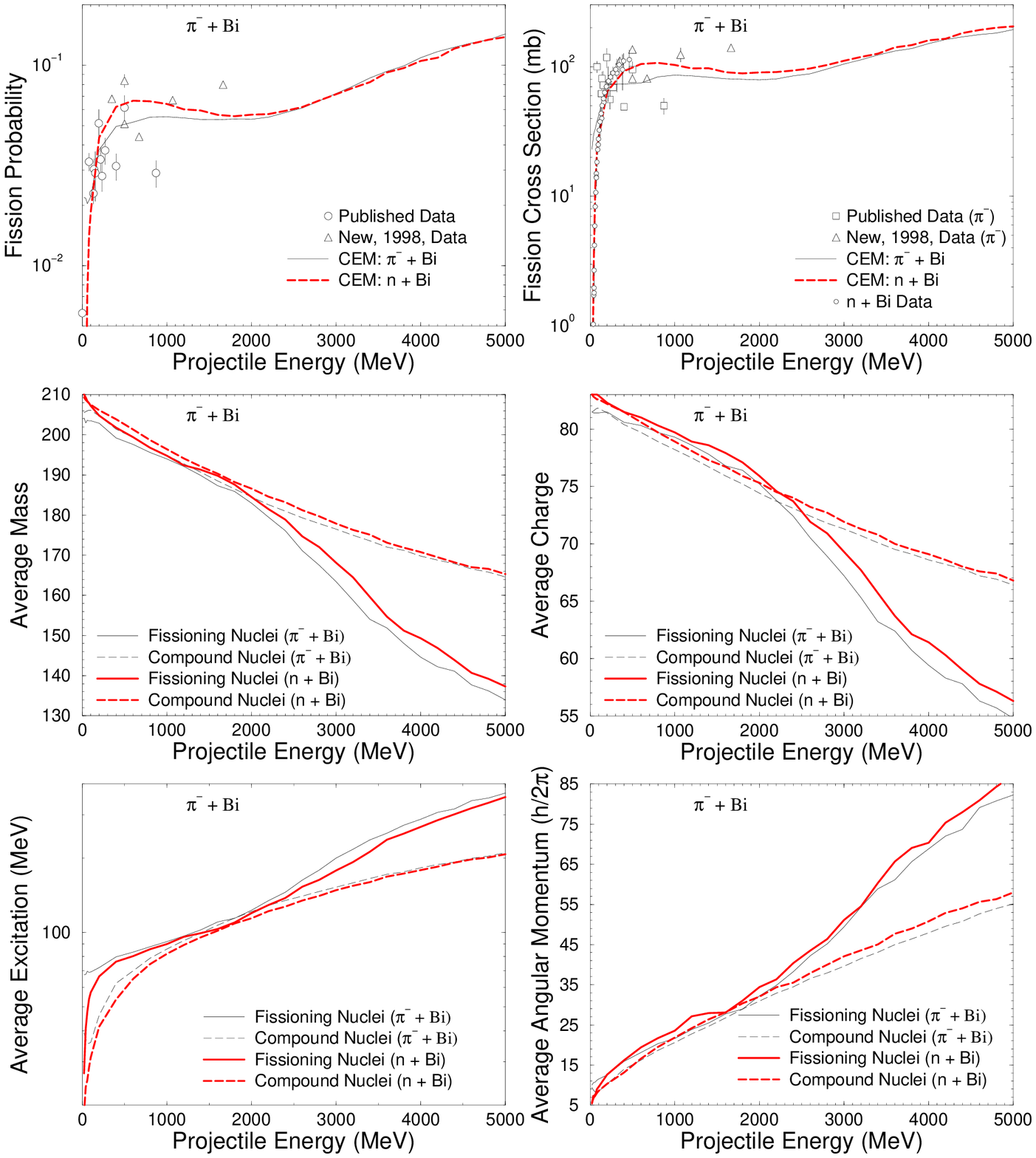,width=175mm,angle=0}}
\end{figure}
\vspace*{0cm}

{\small
Fig.~5.
The energy dependence of neutron- and $\pi^-$-induced fission 
probabilities and fission cross sections of Bi and average values of $Z$,  $A$, 
$E^*$, and $L$ for compound nuclei formed after the preequilibrium stage of 
reactions and for the nuclei which actually do fission, all calculated in CEM95. 
Experimental data for $\pi^-$-induced reactions are from Peterson
\cite{peterson98} and for neutron-induced fission cross sections of Bi from 
the compilation by Prokofiev \cite{prokofiev}.
}

\newpage
\noindent{
al.~\cite{cugnon97}, which has some
success using GEMINI.  In addition, as we see from Fig.~5, 
even the distributions of the compound nuclei remaining after the 
preequilibrium stage of reactions and of those nuclei which actually 
fission, both calculated with the CEM,
are similar only in a restricted range of energy while for a good fraction
of the incident energies from 10 MeV to 5 GeV, these distributions differ
significantly.} 

This means that it is not justified to take just an excited
compound nucleus and to try to adjust all parameters for the fission,
as compound nuclei remaining after the preequilibrium stage of reactions 
in CEM95 and real fissioning nuclei often have quite 
different characteristics, and all calculations 
needed to fix fission parameters have to be done from
the very beginning, making this work very time-consuming.\\

\begin{center}
{\large 4. Summary} \\
\end{center}

We have performed a number of improvements to the Cascade-Exciton model
of nuclear reactions for a better modeling of fission.
These developments include a modified calculation of the level density 
parameter of nuclei at the saddle point,
incorporation into the model of the updated atomic mass tables by 
Audi and Wapstra \cite{audi93}, incorporation of calculated nuclear ground-state 
masses, deformations, and shell corrections by M\"oller et al.~\cite{moller95}, 
and pairing energy shifts from M\"oller, Nix, and Kratz \cite{moller97} 
in level density calculations, derivation of a corrected systematics for level 
density parameters of the Ignatyuk form using the compilation of 
experimental nuclear level density data by Iljinov et al.~\cite{iljinov92} 
and the M\"oller et al.~ground-state microscopic corrections, 
both with and without the M\"oller, Nix, and Kratz pairing gaps,
development of a new empirical relation to take account the thermal 
damping of ground-state shell corrections,
and a number of other refinements to improve the description
of the cascade, preequilibrium, and evaporative stages of reactions,
which also affect calculation of the fission channel.

As we have shown by a number of examples the improvements to the CEM 
made so far clearly have increased its predictive power for fission 
probabilities and cross sections. Our work is not finished.
We hope to find a solution for predicting fission cross sections
for arbitrary targets in a wide range of incident energies 
without introduction of an additional input parameter and we are only
beginning to develop a model appropriate for the CEM to describe mass, 
charge, energy and angular distributions of fission fragments with a possible 
further evaporation of particles from fragments (or even sequential fission) .

\vspace*{0.3cm}
\begin{center}
{\large Acknowledgements} \\
\end{center}
We express our gratitude to R.~J.~Peterson for providing us with part
of his data prior to publication and we thank A.~V.~Prokofiev for his
fruitful collaboration with us on nucleon-induced fission cross sections 
which led to an improvement in the CEM.
We thank N.~V.~Stepanov for providing us with the code of his 
thermodynamical model of fission,
for useful discussions and help in understanding his model 
and we acknowledge J.~R.~Nix and P.~M\"oller for useful discussions. 
We are grateful to R.~E.~MacFarlane and L.~S.~Waters for interest in 
and support of the present work.
This study was supported by the U.~S.~Department of Energy.


\end{document}